\documentstyle[12pt]{article}

\begin{document}

\title{Segre types of symmetric two-tensors \\
       in $n$-dimensional spacetimes}

\author{J. Santos \\
    {\it Universidade Federal do Rio G. do Norte}  \\
    {\it Departamento de F\'{\i}sica, Cx. Postal 1641} \\
    {\it 59072-970 Natal - RN,  Brazil} \\
    {\it Email\/}: janilo@dfte.ufrn.br  \and
         M.J. Rebou\c{c}as and A.F.F. Teixeira \\
    {\it Centro Brasileiro de Pesquisas F\'{\i}sicas}  \\
    {\it Rua Dr.\ Xavier Sigaud, 150}  \\
    {\it 22290-180 Rio de Janeiro - RJ,  Brazil}    \\
    {\it Email\/}: reboucas@cat.cbpf.br  }

\maketitle

\begin{abstract}
Three propositions about Jordan matrices are proved and applied
to algebraically classify the Ricci tensor in $n$-dimensional
Kaluza-Klein-type spacetimes. We show that the possible Segre
types are
$[1,1\ldots 1]$, $[21\ldots 1]$, $[31\ldots 1]$, $[z\bar{z} 1\ldots 1]$
and degeneracies thereof. A set of canonical forms for the Segre types
is obtained in terms of semi-null bases of vectors.
\end{abstract}

{\raggedright
\section{Introduction}  }
\label{intro}

The algebraic classification of a symmetric two-tensor (such as the
Ricci  tensor) defined on a  four-dimensional (4-D for short) Lorentzian
manifold  has been discussed  by several
authors~\cite{Churc} -- \cite{Hall} and is of  interest in,  for example,
classifying and interpreting matter field distributions
\cite{Hall1}~--~\cite{SanRebTei},
and in the study of limits for non-vacuum spacetimes \cite{paiva}.
It is also important in understanding some purely geometrical features of
spacetimes (see, e.g., Pleba\'nski, \cite{Pleban} Cormack and Hall
\cite{CorHal}), and as a part of the procedure for checking whether
apparently different spacetimes are in fact locally the same up to
coordinate transformations (the equivalence
problem~\cite{Karlh}~--~\cite{MM}).

Over the past three decades, particularly after the appearance of
supergravity and superstring theories in the 1970's, there has been a
resurgence in work on Kaluza-Klein-type theories in higher dimensional
settings~\cite{matos}. This has been basically motivated by the  quest
for an unification of gravity with the other fundamental interactions.
On the other hand, they have been  used as a way of finding new
solutions of Einstein's equations  in four dimensions, without ascribing
any physical meaning to  the additional components of the metric
tensor~\cite{Gleiser}.

In this paper, after stating a proposition concerning orthogonality of
null vectors and proving three propositions about Jordan matrices, we
apply them to algebraically classify  the Ricci tensor in $n$-dimensional
Kaluza-Klein-type Lorentzian spacetimes. The  classification is obtained from
first principles, without making use of  previous
results~\cite{Hall,Hall3}, and
generalizes a recent article on Segre types in 5-D spacetimes~\cite{SRT1}.
The Ricci tensor is classified into four Segre types and their
degeneracies. Using real semi-null bases we derive a set of canonical
forms for the Segre types, extending to $n$-dimensional Lorentzian spaces
the canonical forms obtained for lower dimensional
spaces~\cite{Hall}, \cite{Hall3}~--~\cite{SRT2}.
Although the Ricci tensor is  constantly referred to, the results
of the following sections apply to any  second order real symmetric
tensor on $n$-dimensional Lorentzian spaces.

{\raggedright
\section{Mathematical Prerequisites}  }
\label{math}
\setcounter{equation}{0}

The algebraic classification of the Ricci tensor in $n$-dimensional
spacetimes can be cast in terms of the eigenvalue problem
\begin{equation}
\label{eigen}
(R^{a}_{\ b} - \lambda \, \delta^{a}_{b})\,V^{b}\, = 0,
\end{equation}
where $\lambda$ is a scalar, $V^{b}$ is a vector and the mixed
Ricci tensor $R^{a}_{\ b}$ may be thought of as a linear operator
$R: T_{p}(M) \longrightarrow T_{p}(M)$. Here and in what follows
$M$ is a real $n$-dimensional spacetime manifold locally endowed with a
Lorentzian metric of signature $(- + + \cdots +), \; T_{p}(M)$
denotes the tangent space to $M$ at a point $p \in M$ and latin
indices range from $0$ to $n-1$, unless otherwise stated.
Although the matrix $R^a_{\ b}$ is real, the eigenvalues $\lambda$
and the eigenvectors $V^b$ are often complex. A mathematical procedure
used to classify matrices in such a case is to reduce them
through similarity transformations to canonical forms
over the complex field.
Among  the existing canonical forms the Jordan canonical form (JCF)
turns out to be the most appropriate for a classification of
$R^a_{\ b}$.

The mathematical theory concerning JCF of square matrices
is well established and can be found in many textbooks on
linear algebra~\cite{shilov,herstein}. The basic result is that given
an $n$-square matrix $A$ over the complex field, there exist
nonsingular matrices $X$ such that
\begin{equation}
 \label{similar}
    X^{-1}A\,X = J
\end{equation}
where $J$, the JCF of $A$, is a block diagonal matrix, each block
being of the form
\begin{equation}   \label{jblock}
 J_{r}(\lambda_{k})=\left(
 \begin{array}{ccccc}
 \lambda_{k} &    1        &   0    & \cdots & 0 \\
      0      & \lambda_{k} &   1    & \cdots & 0 \\
             &             &        & \ddots &   \\
      0      &    0        &   0    & \cdots & 1 \\
      0      &    0        &   0    & \cdots & \lambda_{k}
  \end{array}
  \right).
\end{equation}
Here $r$ is the dimension of the block and $\lambda _{k}$ is a root
of the characteristic equation  $\mbox{det}(A - \lambda I)=0$. Although
$X$ is far from being unique, the JCF is uniquely determined up to
the ordering of the blocks along the main diagonal of $J$. In what
follows  $A$ of eq.\ (\ref{similar}) is the matrix formed with the
mixed components $R^{a}_{\ b}$ of the Ricci tensor $R$.

We shall first examine the structure of a general Jordan block
$J_{r}(\lambda )$ in $J$, where $J_{r}(\lambda )$ begins at row and column
$s$ and ends at row and column $t$ ($t = s + r - 1$). The matricial
equation (\ref{similar}) can be rewritten as $R\,X = XJ$. Equating
columns $s$ to $t$ on both sides of this matricial equation we have
\begin{equation}
\label{jchain}
  \begin{array}{lrl}
  R\,{\bf X}_{s}     & =  & \lambda \, {\bf X}_{s} \,, \\
  R\,{\bf X}_{s+1}   & =  & \lambda \, {\bf X}_{s+1} + {\bf X}_{s}\,,  \\
                     & \vdots &     \\
  R\,{\bf X}_{t}     & =  & \lambda \, {\bf X}_{t} + {\bf X}_{t-1}\,,
  \end{array}
\end{equation}
where ${\bf X}_{q}$ denotes the vector associated to the $q$-th column
of the matrix $X$.
We shall refer to these relations as a Jordan chain. It should be noted
that (a) the column vectors ${\bf X}_{s},{\bf X}_{s+1}, \ldots ,{\bf X}_{t}$
are linearly independent, otherwise $X$ would be a singular matrix;
(b) the first $m$ vectors in a Jordan chain span an $m$-dimensional
subspace of $T_{p}(M)$ invariant under $R^a_{\ b}$, and (c) particularly
for $m=1$ the vector ${\bf X}_s$ which starts the Jordan chain defines
the sole eigendirection of $R$ associated to that Jordan block.
The complete set of vectors $\{ {\bf X}_{a}\,;\, a=1,\ldots ,n \}$
are linearly independent and give a Jordan basis, i.e., a basis in
which $R^{a}_{\ b}$ takes a JCF.

To algebraically classify a second order {\em symmetric} tensor we
shall discuss now three propositions about Jordan blocks
$J_{r}(\lambda )$ ($r > 1$).
 \newtheorem{guess}{Proposition}
  \begin{guess}
    The first column vector (eigenvector ${\bf X}_{s}$) of a Jordan
    block $J_{r}(\lambda )$ ($ r > 1$) is orthogonal to all vectors
    of its block, except possibly to the last one (${\bf{X}}_t$).
  \end{guess}

Indeed, from  (\ref{jchain}) a vector
${\bf X}_{q+1}$ ($s \leq q < t$) obeys the equation
 \begin{equation}
  \label{vecqua}
  R\,{\bf X}_{q+1} = \lambda {\bf X}_{q+1} + {\bf X}_{q}.
 \end{equation}
Eq.\ (\ref{vecqua}) and the first equation (\ref{jchain}) yield
 \begin{eqnarray}
  \label{innprod}
  \begin{array}{lrl}
  R\,{\bf X}_{q+1}.{\bf X}_{s}  & = & \lambda \, {\bf X}_{q+1}.{\bf X}_{s} +
  {\bf X}_{q}.{\bf X}_{s} \,, \\
  R\,{\bf X}_{s}.{\bf X}_{q+1}  & = & \lambda \, {\bf X}_{s}.{\bf X}_{q+1}\,,
  \end{array}
 \end{eqnarray}
where the dot between vectors indicates inner product. As
$R_{ab}$ is symmetric, one easily obtains
\begin{equation}
    \label{prop1}
     {\bf X}_{s}.{\bf X}_{q} = 0,  \; \; \; s \leq q < t\,.
    \end{equation}
When $q=s$ eq.\ (\ref{prop1}) implies ${\bf X}_{s}.{\bf X}_{s} = 0$, i.e.,
the eigenvector associated to a Jordan block of dimension $r > 1$ is a
null vector.
  \begin{guess}
   If ${\bf X}_{p}$ and ${\bf X}_{q}$ are two vectors (not eigenvectors)
related to a block $J_{r}(\lambda )$ ($r > 2$) then
   \begin{equation}
   \label{prop2}
{\bf X}_{p}.{\bf X}_{q} =
{\bf X}_{p-1}.{\bf X}_{q+1} \,, \; \; \; s <p \leq q < t.
   \end{equation}
  \end{guess}
Indeed, from (\ref{jchain}) one finds that for $p$ and $q$ in the given
range, the  equations
 \begin{eqnarray}
R\,{\bf X}_{p} & = & \lambda\, {\bf X}_{p} + {\bf X}_{p-1}\,,
\label{vecrel1} \\
R\,{\bf X}_{q+1} & = & \lambda\, {\bf X}_{q+1} + {\bf X}_{q} \label{vecrel2}
 \end{eqnarray}
hold.
If one now takes the inner product of (\ref{vecrel1}) and of (\ref{vecrel2})
by ${\bf X}_{q+1}$ and ${\bf X}_{p}$, respectively, and uses the symmetry
of $R_{ab}$, one obtains eqs.\ (\ref{prop2}).
 \begin{guess}
  Eigenvectors related to different Jordan blocks are orthogonal
provided at least one of the blocks has dimension $r > 1$.
 \end{guess}
In fact, let $J_{r}(\lambda )$ and $J_{r'}(\lambda ')$ be blocks
of a Jordan matrix, where $J_{r}(\lambda )$ generates the Jordan chain
(\ref{jchain}) and $J_{r'}(\lambda ')$ gives rise to a similar (primed)
Jordan chain with $r'$ equations. Then
\begin{equation}
 \label{eigenrel}
 R\,{\bf X}_{s'}  = \lambda '{\bf X}_{s'}.
 \end{equation}
{}From (\ref{jchain}) and (\ref{eigenrel}) one finds
 \begin{eqnarray}
R\,{\bf X}_{s'}.{\bf X}_{s}  & = & \lambda'\,{\bf X}_{s'}.{\bf X}_{s}\,,
                                               \label{innrel1}  \\
R\,{\bf X}_{s}.{\bf X}_{s'}  & = & \lambda\, {\bf X}_{s}.{\bf X}_{s'}\,,
                                               \label{innrel2}  \\
R\,{\bf X}_{s+1}.{\bf X}_{s'} & = & \lambda\, {\bf X}_{s+1}.{\bf X}_{s'} +
                             {\bf X}_{s}.{\bf X}_{s'}\,, \label{innrel3} \\
R\,{\bf X}_{s'}.{\bf X}_{s+1} & = & \lambda '\,{\bf X}_{s'}.{\bf X}_{s+1}\,.
                                                 \label{innrel4}
 \end{eqnarray}
Here again the symmetry of $R_{ab}$ together with (\ref{innrel1}) and
(\ref{innrel2}) imply
\begin{equation}
 \label{innrel5}
(\lambda ' - \lambda ){\bf X}_{s'}.{\bf X}_{s} = 0.
\end{equation}
Similarly eqs.\ (\ref{innrel3}) and (\ref{innrel4}) give
\begin{equation}
\label{innrel6}
(\lambda ' - \lambda ){\bf X}_{s'}.{\bf X}_{s+1} = {\bf X}_{s'}.{\bf X}_{s}\,.
\end{equation}
Finally (\ref{innrel5}) and (\ref{innrel6}) imply ${\bf X}_{s'}.{\bf X}_{s}
=0$ regardless of whether $\lambda ' = \lambda $ or $\lambda ' \neq \lambda $.

To complete the mathematical prerequisites we state a trivial extension
of a known result for null vectors in 4-D spacetimes, namely:
\begin{guess}
 In an n-D real vector space endowed with a  Lorentzian metric
two null vectors are orthogonal if and only if they are collinear.
\end{guess}

Employing recurrently proposition 2, then proposition 1,
we have found the null vectors related to a Jordan block $J_{r}(\lambda )$
with $r = 2, 3, 4$. These null vectors and their orthogonality
relations with other vectors of the same block are
shown in table \ref{nullvectors}. Although we have included in table
\ref{nullvectors} the null vectors when $r=4$,
it should be noticed that in the classification of symmetric two-tensors
in $n$-dimensional Lorentzian spaces the Jordan blocks of dimension $r \geq 4$
cannot occur.
In the next section we shall discuss this point, and use it together with
propositions 3 and 4 to classify $R^a_{\ b}$.
\begin{table}
\caption{Null vectors related to a Jordan block $J_r(\lambda )$.}
\label{nullvectors}
\begin{center}
\begin{tabular}{|c|l|l|} \hline
$r$ & null vectors & orthogonality relations  \\  \hline
\hline
2 & ${\bf X}_1$               &                           \\   \hline
3 & ${\bf X}_1$               & ${\bf X}_1.{\bf X}_2 = 0$  \\  \hline
4 & ${\bf X}_1$, ${\bf X}_2$  & ${\bf X}_1.{\bf X}_2 =
                                {\bf X}_1.{\bf X}_3 = 0$  \\  \hline
\end{tabular}
\end{center}
\end{table}

\vspace{2mm}
{\raggedright
\section{The Classification}  }
\label{class}
\setcounter{equation}{0}

In the Jordan classification two square matrices are
said to be equivalent if similarity transformations exist
such that they can be brought to the same JCF. The JCF of
a matrix gives explicitly its eigenvalues and makes apparent
the dimensions of the Jordan blocks.
However, for many purposes a somehow coarser classification
of a matrix is sufficient. In the Segre classification,
for example, the value of the roots of the characteristic
equation is not relevant --- only dimension of the Jordan
blocks and degeneracy of eigenvalues matter.
The Segre type is a list $[r_1 r_2 \cdots r_m]$ of the
dimensions of the Jordan blocks. Equal eigenvalues in distinct
blocks are indicated by enclosing the corresponding digits
inside round brackets. Thus, for example, in the degenerated Segre
type $[(42)1]$ six out of the seven eigenvalues are equal;
there are three linearly independent eigenvectors, two of which
are associated to the Jordan blocks of dimensions 4 and 2, whereas
the last one corresponds to the block of dimension 1.

In classifying symmetric tensors in a Lorentzian spacetime two
refinements to the usual Segre notation are often used.
Instead of a digit to denote the dimension of a block with
complex eigenvalue a letter is used, and the digit corresponding
to a timelike eigenvector is separated from the others
by a comma.

We learn from table \ref{nullvectors} that the JCF of $R^{a}_{\ b}$
cannot have a block with dimension greater than 3. Indeed, the
existence of such a block would give rise to at least two
null vectors ${\bf X}_{1}$ and ${\bf X}_{2}$ that are orthogonal,
hence collinear by proposition 4, which would lead to a singular
transformation matrix $X$.
Besides, the JCF of $R^{a}_{\ b}$ cannot have more than one block of
dimension $r > 1$. This is so because there is one null eigenvector
associated to each such block and, by proposition 3, these vectors
are orthogonal, hence the matrix $X$ in (\ref{similar}) would again be
singular.
Therefore, in a 5-D spacetime, for example, the
JCF with Segre types $[5]$, $[41]$, $[32]$, $[221]$ and their
degeneracies are ruled out by the above arguments, in agreement
with earlier results on this matter~\cite{SRT1}.
Similarly, one can find the possible non-degenerated Segre types
of  $R^{a}_{\ b}$ for $n$-dimensional ($n \geq 3$) spacetimes. In table
\ref{segretypes} we present these types, where we have included
the complex cases to be discussed below.
\begin{table}
\caption{Segre types for $R^{a}_{\ b}$ in $n$-dimensional spacetimes.}
\label{segretypes}
\begin{center}
\begin{tabular}{|c|l|} \hline
dimension of &  \\
spacetime & Segre types (non-degenerated)  \\  \hline \hline
3 & $[1,11]$, $\; [21]$, $\; [3]$,  $\; [z\bar{z} 1]$  \\ \hline
4 & $[1,111]$, $\; [211]$, $\; [31]$, $\; [z\bar{z} 11]$  \\ \hline
5 & $[1,1111]$, $\; [2111]$, $\; [311]$, $\; [z\bar{z} 111]$  \\ \hline
$\vdots $ & $\vdots $  \\  \hline
$n$ & $[1,1\ldots 1]$, $\; [21\ldots 1]$, $\; [31\ldots 1]$,
$\; [z\bar{z}11\ldots 1]$  \\ \hline
\end{tabular}
\end{center}
\end{table}

When the characteristic equation corresponding to (\ref{eigen}) has complex
roots, one can deal with this case by using an approach borrowed
{}from~\cite{Hall} as follows. Suppose that
$\alpha \pm i\beta $ are complex eigenvalues of $R^{a}_{\ b}$
corresponding to  the eigenvectors
${\bf V}_{\pm} = {\bf Y}\pm i{\bf Z}$, where  $\alpha$ and $\beta \neq 0$
are real and ${\bf Y}, \, {\bf Z}$  are independent vectors defined on
$T_{p}(M)$. Since  $R_{ab}$ is symmetric and the eigenvalues are different,
the eigenvectors  must be orthogonal and hence equation
${\bf Y}.{\bf Y} + {\bf Z}.{\bf Z} = 0$ holds.
It follows that either one of the vectors ${\bf Y}$ or ${\bf Z}$ is
timelike and the other spacelike or both are null and,
since $\beta \neq 0$, not collinear. Whichever is the
case, the real and the imaginary part of (\ref{eigen}) give
\begin{eqnarray}
R^{a}_{\ b}\,Y^{b} & = & \alpha Y^{a} - \beta Z^{a}\,,
\label{relc1}  \\
R^{a}_{\ b}\,Z^{b} & = & \beta Y^{a} + \alpha Z^{a}\,.  \label{relc2}
\end{eqnarray}
Thus, the vectors ${\bf Y}$ and ${\bf Z}$ span a timelike
two-dimensional subspace  of $T_p(M)$ invariant under
$R^{a}_{\ b}\,$. Besides, by a procedure similar to that used
in 5-D Lorentzian spaces~\cite{SRT1} one can show that the
($n-2$)-dimensional
space orthogonal to this timelike 2-space is spacelike,
is also invariant under $R^{a}_{\ b}$ and contains $n-2$
orthogonal eigenvectors of $R^{a}_{\ b}$ with real eigenvalues.
These eigenvectors, together with ${\bf V}_{+}$ and ${\bf V}_{-}$,
form a set of $n$ linearly independent eigenvectors of $R^{a}_{\ b}$
at $p \in M$. Therefore, when there exists complex (a conjugate pair of)
eigenvalues, $R^{a}_{\ b}$ is necessarily diagonalizable over the
complex field and possesses $n-2$ real eigenvalues. Its Segre type
is $[z\bar{z}1\ldots 1]$ or one of its degeneracies.

\vspace{2mm}

{\raggedright
\section{A Set of Canonical Forms} }
\label{base}
\setcounter{equation}{0}

An often used approach in physics to establish a canonical form
for a tensor is to align the basis vectors along the preferred
directions intrinsically defined by the tensor.
As far as the Ricci tensor $R$ is concerned, the existence of
null eigenvectors suggests to choose a semi-null basis $\cal B$
for $T_{p}(M)$, consisting of 2 null vectors and $n-2$ spacelike
vectors,
\begin{equation}
\label{basis}
{\cal B} =\{{\bf l},{\bf m},{\bf x}^{(1)},{\bf x}^{(2)},\ldots ,
{\bf x}^{(n-2)}\}\,,
\end{equation}
such that the only non-vanishing inner products are
\begin{equation} \label{inerp}
{\bf l}.{\bf m}={\bf x}^{(1)}.{\bf x}^{(1)}={\bf x}^{(2)}.{\bf x}^{(2)}=
\ldots ={\bf x}^{(n-2)}.{\bf x}^{(n-2)}=1
\end{equation}
The most general decomposition of $R_{ab}$ in terms of the
basis $\cal B$ is
\begin{eqnarray}
\label{rabgen}
R_{ab} & = & 2\rho_{1}l_{(a}m_{b)}+\rho_{2}l_{a}l_{b} + \nonumber \\
 & & \rho_{3}x^{(1)}_{a}x^{(1)}_{b} + \rho_{4}x^{(2)}_{a}x^{(2)}_{b}
+ \cdots +
\rho_{n}x^{(n-2)}_{a}x^{(n-2)}_{b} + \rho_{n+1}m_{a}m_{b} + \nonumber \\
 & & 2\rho_{n+2}l_{(a}x^{(1)}_{b)} + \cdots
     + 2\rho_{n(n+1)/2}x^{(n-3)}_{(a}x^{(n-2)}_{b)}\,,
\end{eqnarray}
where the coefficients $\rho_{1}, \ldots ,\rho_{n(n+1)/2}$  are real scalars.

We shall now show that semi-null bases $\cal B$ can always be chosen so
that $R_{ab}$ takes one of the following canonical forms at $p \in M$:
\begin{eqnarray}
\mbox{\bf Segre type} & &  \mbox{\bf Canonical form} \nonumber \\
{[}1,1\ldots 1] & R_{ab} = & 2\rho_{1}l_{(a}m_{b)} + \rho_{2}(l_{a}l_{b} +
m_{a}m_{b}) + \rho_{3}x^{(1)}_{a}x^{(1)}_{b} + \rho_{4}x^{(2)}_{a}x^{(2)}_{b}
 \nonumber  \\
 & & + \cdots + \rho_{n}x^{(n-2)}_{a}x^{(n-2)}_{b}\,, \label{rab11111}  \\
{[}21\ldots 1] & R_{ab} = & 2\rho_{1}l_{(a}m_{b)} \pm l_{a}l_{b} +
\rho_{3}x^{(1)}_{a}x^{(1)}_{b} + \rho_{4}x^{(2)}_{a}x^{(2)}_{b}
 \nonumber  \\
 & & + \cdots + \rho_{n}x^{(n-2)}_{a}x^{(n-2)}_{b}\,, \label{rab2111}  \\
{[}31\ldots 1] & R_{ab} = & 2\rho_{1}l_{(a}m_{b)} + 2l_{(a}x_{b)} +
\rho_{1}x^{(1)}_{a}x^{(1)}_{b} + \rho_{4}x^{(2)}_{a}x^{(2)}_{b}
 \nonumber  \\
 & & + \cdots + \rho_{n}x^{(n-2)}_{a}x^{(n-2)}_{b}\,,  \label{rab311}  \\
{[}z\bar{z}11\ldots 1] & R_{ab} = & 2\rho_{1}l_{(a}m_{b)} +
\rho_{2}(l_{a}l_{b} -
m_{a}m_{b}) + \rho_{3}x^{(1)}_{a}x^{(1)}_{b} + \rho_{4}x^{(2)}_{a}x^{(2)}_{b}
 \nonumber  \\
 & & + \cdots + \rho_{n}x^{(n-2)}_{a}x^{(n-2)}_{b}\,, \label{rabZZ}
\end{eqnarray}
where the coefficients $\rho_{1}, \ldots ,\rho_{n}$  are real scalars and
$\rho_{2}\neq 0$ in (\ref{rabZZ}). Clearly these coefficients are related
to the eigenvalues of $R^a_{\ b}$.

Using the non-vanishing inner products of the basis $\cal B$, it is
not difficult to show that each of the above expressions for
$R_{ab}$ leads to the corresponding Segre type indicated on the left.
However, to show the reciprocal is not as simple as that and we shall
examine case by case.

{\bf Segre type [1,1\ldots 1]}. For this Segre type one writes down a
$n$-dimensional general real symmetric matrix for the metric tensor
$g_{ab}$ and imposes the condition
\begin{equation}
\label{restric}
g_{ac}\,R^{c}_{\ b} =
g_{bc}\,R^{c}_{\ a}\,
\end{equation}
to account for the symmetry of $R_{ab}$. In a basis  where
$R^{a}_{\ b}$ is diagonal (\ref{restric}) implies that $g_{ab}$
is also diagonal. As $\mbox{det}\:(g_{ab}) < 0$, then all
$g_{ii} \neq 0$. Each basis vector is an eigenvector of $R^{a}_{\ b}$
and has norm $g_{ii}$, so they are either timelike
or spacelike and orthogonal to each other. Actually,
owing to the Lorentzian signature we have one timelike
and $n-1$ spacelike vectors, which suitably normalized give
a $n$-dimensional Minkowski basis
$\bar{\cal B} = \{{\bf t},{\bf w},{\bf x}^{(1)},{\bf x}^{(2)},\ldots ,
{\bf x}^{(n-2)}\}$ where ${\bf t}$ is the timelike eigenvector.
If one now writes down the expression for $R_{ab}$ in terms of this
Minkowski basis, and then introduces the null vectors
${\bf l} = \frac{1}{\sqrt{2}} ({\bf w} + {\bf t})$ and
${\bf m} = \frac{1}{\sqrt{2}} ({\bf w} - {\bf t})$
to form a semi-null basis
${\cal B}= \{{\bf l},{\bf m},{\bf x}^{(1)},{\bf x}^{(2)},\ldots ,
{\bf x}^{(n-2)}\}$, one finally obtains $R_{ab}$ in the canonical
form (\ref{rab11111}). The eigenvalues are
$\rho_{1} - \rho_{2},\, \rho_{1} + \rho_{2},\, \rho_{3},\,
\rho_{4}, \ldots , \rho_{n}$ and the
corresponding eigenvectors are ${\bf l} - {\bf m},\; {\bf l} + {\bf m},\;
{\bf x}^{(1)},\; {\bf x}^{(2)},\, \ldots , {\bf x}^{(n-2)}$.

{\bf Segre type [21\ldots 1]}.
{}From table \ref{nullvectors} one learns that for this case
the first vector of the Jordan basis, namely the eigenvector
${\bf X}_1$, is a null vector,
and using proposition 3 one finds that the eigenvectors
${\bf X}_{3},{\bf X}_{4},\ldots ,{\bf X}_{n}$ are spacelike,
mutually orthogonal and orthogonal to ${\bf X}_1$.
We are then naturally led to choose ${\bf l}$ along the direction of
${\bf X}_1$, and ${\bf x}^{(1)},\ldots ,{\bf x}^{(n-2)}\,$ along the
directions of ${\bf X}_{3},\ldots ,{\bf X}_{n}$, respectively.
To complete the semi-null basis we choose ${\bf m}$ along the
direction uniquely defined by
${\bf l},{\bf x}^{(1)},\ldots ,{\bf x}^{(n-2)}\,$. Imposing that
these last vectors are eigenvectors of $R$
one finds that $R_{ab}$ in (\ref{rabgen})
simplifies to
\begin{eqnarray}
\label{rab211}
R_{ab} & = & 2\rho_{1}l_{(a}m_{b)} + \rho_{2}l_{a}l_{b} +
\rho_{3}x^{(1)}_{a}x^{(1)}_{b} + \rho_{4}x^{(2)}_{a}x^{(2)}_{b}
 \nonumber  \\
 & & + \cdots + \rho_{n}x^{(n-2)}_{a}x^{(n-2)}_{b}\,,
\end{eqnarray}
where the condition $\rho_{2} \neq 0$ must be imposed otherwise
${\bf m}$ would be a $n$-th linearly independent eigenvector.
We finally make the transformation
${\bf l} \rightarrow {\bf l} \,{\left|\rho_{2} \right|}^{-1/2}\,,$
${\bf m} \rightarrow {\bf m} \,{\left|\rho_{2} \right|}^{1/2}\,,$
to bring $R_{ab}$ from the general form (\ref{rabgen})
into the form (\ref{rab2111}) with
$\rho_{1},\, \rho_{3},\, \ldots , \rho_{n}$ the eigenvalues associated to
the eigenvectors
${\bf l},\;{\bf x}^{(1)},\; \ldots , {\bf x}^{(n-2)}$.

{\bf Segre type [31\ldots 1]}.
For this Segre type one learns from table \ref{nullvectors}
that the Jordan basis contains a null vector ${\bf X}_1$,
orthogonal to the spacelike vector ${\bf X}_2$. Moreover,
employing proposition 3 one also finds that
${\bf X}_{4},{\bf X}_{5},\ldots ,{\bf X}_{n}$ are spacelike
eigenvectors orthogonal to ${\bf X}_1$, and a further inspection
of the various Jordan chains shows that they are also mutually
orthogonal as well as orthogonal to ${\bf X}_2$ and ${\bf X}_3$.
Nevertheless the vectorial character of ${\bf X}_3$ remains
open. So, to form a semi-null basis for this case
a new basis $\{ {\bf \tilde{X}}_a; a=1, \cdots ,n \}$
ought to be found. It can be shown that the following basis
transformation leaves invariant the matrix $J^a_{\ b}$:
\begin{equation}
\label{newbasis}
  \begin{array}{lrl}
{\bf \tilde{X}}_1 & = & a {\bf X}_{1} \,, \\
{\bf \tilde{X}}_2 & = & a {\bf X}_{2} + b {\bf X}_{1} \,, \\
{\bf \tilde{X}}_3 & = & a {\bf X}_{3} + b {\bf X}_{2} + c {\bf X}_{1} \,, \\
{\bf \tilde{X}}_i & = & d_i {\bf X}_{i} \; \qquad \mbox{(no sum)}\,,
  \end{array}
\end{equation}
where $a, b, c, d_i$ ($i=4, \cdots ,n$) are $n$ arbitrary real
constants with $a \neq 0$ and $d_i \neq 0$. To endow the new
basis $\{ {\bf \tilde{X}}_a \}$ with the orthonormality
relations of a semi-null basis (\ref{inerp})
we simply have to impose the $n$ constraints
\begin{equation}
{\bf \tilde{X}}_2.{\bf \tilde{X}}_2 =
{\bf \tilde{X}}_4.{\bf \tilde{X}}_4 =
\cdots =
{\bf \tilde{X}}_n.{\bf \tilde{X}}_n = 1 \,,
\end{equation}
\begin{equation}
{\bf \tilde{X}}_2.{\bf \tilde{X}}_3 =
{\bf \tilde{X}}_3.{\bf \tilde{X}}_3 = 0\,.
\end{equation}
The constraints define the values of $a, b, c, d_i$
in (\ref{newbasis}), namely
\begin{equation}
a=({\bf X}_2.{\bf X}_2)^{-1/2}\,, \qquad \qquad \quad b=- \frac{1}{2} a^3
{\bf X}_2.{\bf X}_3\,,
\end{equation}
\begin{equation}
c= \frac{3}{2} b^2 a^{-1} - \frac{1}{2} a^3 {\bf X}_3.{\bf X}_3,
\qquad
d_i = ({\bf X}_i.{\bf X}_i)^{-1/2}\,.
\end{equation}
We finally choose ${\bf l} = {\bf \tilde{X} }_1,\,\, {\bf x}^{(1)} =
{\bf \tilde{X} }_2,\,\, {\bf m} = {\bf \tilde{X} }_3,\,\, {\bf x}^{(i-2)} =
{\bf \tilde{X} }_i$  ($i = 4, \cdots ,n$) and write
$R_{ab}$ as (\ref{rab311}) with $\rho_1, \rho_4, \rho_5, \cdots ,\rho_n$
the eigenvalues associated to the eigenvectors
${\bf l},{\bf x}^{(2)},\ldots ,{\bf x}^{(n-2)}$,
respectively.

{\bf Segre type ${\bf [z\,\bar{z}1\ldots 1]}$}.
For this type all vectors of a Jordan basis are eigenvectors
of $R^a_{\ b}$. Two of them, namely
${\bf V}_{\pm}  = {\bf Y} \pm i\, {\bf Z}$, are complex
conjugate with eigenvalues $ \alpha \pm i\, \beta$, respectively
(see section 2). The other basis vectors
${\bf X}_3, \cdots ,{\bf X}_n$ are spacelike eigenvectors
mutually orthogonal and orthogonal to the timelike 2-D
space spanned by ${\bf Y}$ and ${\bf Z}$. Through a linear
combination of ${\bf Y}$ and ${\bf Z}$ two null vectors
${\bf l}$ and ${\bf m}$ can be constructed satisfying
${\bf l}.{\bf m} = 1$ and
\begin{eqnarray}
R^{a}_{\ b}\,l^{b} & = & \alpha l^{a} - \beta m^{a}\,,
\label{rel1}  \\
R^{a}_{\ b}\,m^{b} & = & \beta l^{a} + \alpha m^{a}\,.  \label{rel2}
\end{eqnarray}
One can then form a semi-null basis
$\{{\bf l},{\bf m},{\bf x}^{(1)},{\bf x}^{(2)},\ldots ,
{\bf x}^{(n-2)}\}$,
where each ${\bf x}^{(i)}$ is the spacelike eigenvector  ${\bf X}_i$
suitably normalized. In terms of this basis, and taking into
account (\ref{rel1}) and (\ref{rel2}), the Ricci tensor
reduces from the general form ({\ref{rabgen}) to the form (\ref{rabZZ}),
where $\rho_1 = \alpha$ and $\rho_2 = \beta$. The eigenvalues
are $\rho_1+i \rho_2,\, \rho_1 - i \rho_2,\, \rho_3, \cdots ,\rho_n$ and the
corresponding eigenvectors are
$ {\bf l} + i {\bf m}\,, {\bf l} - i {\bf m},\, {\bf x}^{(1)}, \cdots
,{\bf x}^{(n-2)}$.
\vspace{2mm}

{\raggedright
\section{Acknowledgments} }

We take this occasion to express our warmest thanks to our friend
Graham Hall for many motivating and valuable discussions during
his visit to CBPF (Brazilian Center for Physics Research).

\end{document}